\documentclass[english]{article}

\usepackage{amssymb}
\usepackage{amsfonts}
\usepackage{amsmath}


\begin{document}

\title{In- and out-states of scalar particles confined between two capacitor
plates}

\author{A.I. Breev$^{1}$\thanks{
breev@mail.tsu.ru}, S.P. Gavrilov$^{1,2}$\thanks{
gavrilovsergeyp@yahoo.com; gavrilovsp@herzen.spb.ru}, and D.M. Gitman$
^{1,3,4}$\thanks{
gitman@if.usp.br} \\
{\normalsize $^{1}$ Department of Physics, Tomsk State University, Tomsk
634050, Russia.}\\
{\normalsize $^{2}$ Department of General and Experimental Physics, }\\
{\normalsize Herzen State Pedagogical University of Russia,}\\
{\normalsize Moyka embankment 48, 191186 St. Petersburg, Russia;}\\
{\normalsize $^{3}$ P.N. Lebedev Physical Institute, 53 Leninsky prospekt,
119991 Moscow, Russia;}\\
{\normalsize $^{4}$ Institute of Physics, University of S\~{a}o Paulo, CEP
05508-090, S\~{a}o Paulo, SP, Brazil; }}
\maketitle

\begin{abstract}
In the present article, using a non-commutative integration method of linear
differential equations, we, considering the Klein-Gordon equation with the 
$L$-constant electric field with large $L$ and using the light cone
variables, find new complete sets of its exact solutions. These solutions
can be related by integral transformations to previously known solutions
that were found in Phys. Rev. D. \textbf{93}, 045033(2016). Then, using the
general theory developed in Phys. Rev. D. \textbf{93}, 045002 (2016), we
construct (in terms of the new solutions) the so-called \textrm{in}- and 
\textrm{out}-states of scalar particles confined between two capacitor
plates.
\end{abstract}

\section{Introduction\label{S1}}

A particle production from a vacuum by strong electric-like external
backgrounds (the Schwinger effect \cite{Schwinger51}, or the effect of the
vacuum instability) is one of the most interesting effect in quantum field
theory (QFT) that attracts attention already for a long time. The effect can
be observable if the external fields are sufficiently strong, e.g. the
magnitude of an electric field should be comparable with the Schwinger
critical field $E_{\mathrm{c}}=m^{2}c^{3}/e\hslash \simeq 10^{16}\mathrm{V/cm
}$. Nevertheless, recent progress in laser physics allows one to hope that
an experimental observation of the effect can be possible in the near
future, see Refs. \cite{Dun09} for a review. Moreover, electron-hole pair
creation from the vacuum becomes also observable in laboratory conditions in
the graphene and similar nanostructures, see, e.g. Refs. \cite{dassarma}.
Depending on the structure of such external backgrounds, different
approaches have been proposed for calculating the effect, a list of relevant
publications can be found in Refs. \cite{RufVSh10,GelTan16}. Calculating
quantum effects in strong external backgrounds must be nonperturbative with
respect to the interaction with strong backgrounds. A general formulation of
QED with time-dependent external fields (so-called $t$-potential steps) was
developed in Refs. \cite{Gitman}. It can be also seen that in some
situations in graphene and similar nanostructures the vacuum instability
effects caused by strong (with respect of massless fermions) electric fields
are of significant interest; see, e.g., Refs. \cite{GelTan16,allor08,GavGitY12,
VafVish14,KaneLM15,Olad+etal17,Akal+etal19} and
references therein. At the same time, in these cases electric fields can be
considered as time-independent weakly inhomogeneous $x$-electric potential
steps (electric fields of constant direction that are concentrated in
restricted space areas) that can be approximated by a linear potential.
Approaches for treating quantum effects in the explicitly time-dependent
external fields are not directly applicable to the $x$-electric potential
steps. In the recent work \cite{x-case} a consistent nonperturbative
formulation of QED with critical $x$-electric potential steps, strong enough
to violate the vacuum stability, was constructed. There a nonperturbative
calculation technique for different quantum processes such as scattering,
reflection, and electron-positron pair creation was developed. This
technique essentially uses special sets of exact solutions of the Dirac and
Klein Gordon equation with the corresponding external field of $x$-electric
potential steps. The cases when such solutions can be found explicitly
(analytically) are called exactly solvable cases. This technique was
effectively used to describe particle creation effect in the Sauter field of
the form $E(x)=E\cosh ^{-2}\left( x/L_{\mathrm{S}}\right)$, in a constant
electric field between two capacitor plates separated by a distance $L$ (the
so-called $L$-constant electric field), and in exponential time-independent
electric steps, where the corresponding exact solutions were available, see
Refs. \cite{x-case,L-field,x-exp}. These exactly solvable models allowed one
to develop a new approximate calculation method to treat nonperturbatively
the vacuum instability in arbitrary weakly-inhomogeneous $x$-electric
potential steps \cite{GGSh19}. Note, that the corresponding limiting case of
a constant uniform electric field has many similarities with the case of the
de Sitter background, see, e.g., Refs.~\cite{AndMot14,AkhmP15} and
references therein. Thus, a study of the vacuum instability in the presence
of the $L$-constant electric field with large $L\rightarrow \infty$ may be
quite important for some applications. Only critical step with a potential
difference $\Delta U > 2m$ ($m$ is the electron mass) can produce
electron-positron pairs, moreover, pairs are born only with quantum numbers
in a finite range, in the so-called Klein zone.

As a matter of fact, non-perturbative calculation techniques are related to
the possibility of constructing exact solutions of the corresponding
relativistic Dirac or Klein Gordon equations, in particular, solutions that
have special asymptotics. Constructing of such solutions is a rather
difficult task. Sometimes, to solve it, an adequate choice of variables in
the corresponding equations is useful. In particular, in the work 
\cite{NaNi76}, see \cite{GGG98} as well, considering the Dirac or Klein Gordon
equations with a constant uniform field given by time-dependent potential
and choosing the variables of the light cone, the above solutions in a
special representation were found. With these solutions, it was possible to
find explicitly all kinds of the corresponding QED singular functions in the
Fock-Schwinger proper time representation. In the present article, using a
non-commutative integration method of linear differential equations, we,
considering the Klein-Gordon equation with the $L$-constant electric field
with large $L$ and using the light cone variables, find new complete sets of
its nonstationary exact solutions. These solutions can be related by
integral transformations to previously known stationary solutions that were
found in Ref. \cite{L-field}. Then, using the general theory developed in
Ref. \cite{x-case}, we construct (in terms of the new nonstationary
solutions) the so-called \textrm{in}- and \textrm{out}-states of scalar
particles confined between two capacitor plates.

\section{In- and out-solutions\label{S2}}

Here we construct \textrm{in}- and \textrm{out}-solutions of the
Klein-Gordon equation with an external constant electric field, which is the
so-called $L$-constant electric field and belongs to the class of $x$
-potential steps. The equation has the form: 
\begin{eqnarray}
&&\left( P^{\mu }P_{\mu }-m\right) \psi (X)=0,\ P_{\mu }=i\partial _{\mu
}-qA_{\mu }(X),  \notag \\
&&P^{\mu }=\eta ^{\mu \nu }P_{\nu },\ \eta ^{\mu \nu }=\mathrm{diag}\underset
{d}{\underbrace{\left( 1,-1,\dots ,-1\right) }},\ d=D+1,  \label{1.1}
\end{eqnarray}
where $A_{\mu }(X)$ are corresponding electromagnetic potentials, $m$ is the
particle mass and $q=-e$, $e>0$ is its charge. For generality, we consider
the problem in $d$-dimensional spacetime ($\hbar =c=1$). Here $X=(X^{\mu
})=(t,\mathbf{r})$, $\mathbf{r}=(X^{k})$, $\mu =0,1,\dots D$, $k=1,\dots ,D$
. The $L$-constant electric field $E\left( x\right) $ has the form 
\begin{equation}
E\left( x\right) =\left\{ 
\begin{array}{ll}
0, & x\in (-\infty ,-L/2]\cup \lbrack L/2,\infty ) \\ 
E, & x\in (-L/2,L/2)
\end{array}
\right. ,\quad L>0.  \label{g1}
\end{equation}

We assume that the corresponding $x$-potential step is critical and
sufficiently large, such that $eEL\gg 2m$. In this case the field $E(x)$ and
leading contributions to vacuum mean values can be considered as macroscopic
ones. At the same time, this $L$-constant electric field is weakly
inhomogeneous, the corresponding Klein zone is extensive, such that all the
universal properties of the vacuum instability described in Ref. \cite
{GGSh19} hold true. In the limit $L\rightarrow \infty$ the $L$ -constant
field is a kind of a regularization for a constant uniform electric field.
In fact, in this limit we may approximate the $L$-constant field by a
constant uniform electric field given by a linear potential, 
\begin{equation}
A_{0}(X)=-Ex,\quad A_{k}(X)=0,\quad x=X^{1},\quad E>0.  \label{g2}
\end{equation}

Let us consider stationary solutions of the Klein-Gordon equation, having
the following form: 
\begin{eqnarray}
&&\psi _{n}\left( X\right) =\varphi _{n}\left( t,x\right) \varphi _{\mathbf{p
}_{\bot }}\left( \mathbf{r}_{\bot }\right) ,\quad \varphi _{\mathbf{p}_{\bot
}}\left( \mathbf{r}_{\bot }\right) =(2\pi )^{-(d-1)/2}\,\exp \left( i\mathbf{
p}_{\bot }\mathbf{r}_{\bot }\right) ,\quad X=\left( t,x,\mathbf{r}_{\bot
}\right) ,  \notag \\
&&\varphi _{n}\left( t,x\right) =\exp \left( -ip_{0}t\right) \varphi
_{n}\left( x\right) ,\;\ n=(p_{0},\mathbf{p}_{\bot }),  \notag \\
&&\mathbf{r}_{\bot }=\left( X^{2},\ldots ,X^{D}\right) ,\ \mathbf{p}_{\bot
}=\left( p^{2},\ldots ,p^{D}\right) ,\ \hat{p}_{x}=-i\partial _{x}.
\label{e2}
\end{eqnarray}%
These solutions are quantum states of spinless particles with given energy $
p_{0}$ and momenta $\mathbf{p}_{\bot }$ perpendicular to the $x$-direction.
The functions $\varphi _{n}\left( x\right) $ obey the second-order
differential equation 
\begin{equation}
\left\{ \hat{p}_{x}^{2}-\left[ p_{0}-U\left( x\right) \right] ^{2}+\mathbf{p}
_{\bot }^{2}+m^{2}\right\} \varphi _{n}\left( x\right) =0,\quad
U(x)=-eA_{0}(x).  \label{e3}
\end{equation}

We would like to construct two complete sets of solutions of form (\ref{e2}
), we denote them as $_{\;\zeta }\psi _{n}\left( X\right)$ and $
^{\;\zeta}\psi _{n}\left( X\right)$, $\zeta =\pm$ in what follows, with
special left and right asymptotics, 
\begin{eqnarray*}
&&\hat{p}_{x}\ _{\zeta }\psi _{n}\left( X\right) =p^{\mathrm{L}}\ _{\;\zeta
}\psi _{n}\left( X\right) ,\ \ x\rightarrow -\infty ,  \notag \\
&&\hat{p}_{x}\ ^{\zeta }\psi _{n}\left( X\right) =p^{\mathrm{R}}\ ^{\zeta
}\psi _{n}\left( X\right) ,\ \ x\rightarrow +\infty.
\end{eqnarray*}

The solutions $_{\;\zeta }\psi _{n}\left( X\right) $ and $^{\;\zeta }\psi
_{n}\left( X\right) $ asymptotically describe particles with given real
momenta $p^{\mathrm{L/R}}$ along the $x$ direction. The corresponding
functions $\varphi _{n}\left( x\right) $ are denoted by $_{\;\zeta }\varphi
_{n}\left( x\right) $ and $^{\;\zeta }\varphi _{n}\left( x\right)$
respectively. These functions have the asymptotics: 
\begin{eqnarray*}
&&_{\zeta }\varphi _{n}\left( x\right) =\,_{\zeta }\mathcal{N}~\exp \left[
i\left\vert p^{\mathrm{L}}\right\vert x\right] ,\quad x\rightarrow -\infty ,
\\
&&^{\zeta }\varphi _{n}\left( x\right) =\,^{\zeta }\mathcal{N}\,\exp \left[
i\left\vert p^{\mathrm{R}}\right\vert x\right] ,\quad x\rightarrow +\infty .
\end{eqnarray*}

Solutions $_{\zeta }\psi _{n}\left( X\right) $ and $^{\zeta }\psi _{n}\left(
X\right) $ are subjected to the following orthonormality conditions with
respect to the Klein-Gordon inner product on the $x=\mathrm{const}$
hyperplane:
\begin{eqnarray}
\left( \ _{\zeta }\psi _{n},\ _{\zeta ^{\prime }}\psi _{n^{\prime }}\right)
_{x} &=&\left( \ ^{\zeta }\psi _{n},\ ^{\zeta ^{\prime }}\psi _{n^{\prime
}}\right) _{x}=\zeta \delta _{\zeta ,\zeta ^{\prime }}\delta _{n,n^{\prime
}},  \notag \\
\delta _{n,n^{\prime }} &=&\delta \left( p_{0}-p_{0}^{\prime }\right) \delta
\left( \mathbf{p}_{\bot }-\mathbf{p}_{\bot }^{\prime }\right) ,  \notag \\
\left( \psi ,\psi ^{\prime }\right) _{x} &=&i\int \psi ^{\ast }\left(
X\right) \left( \overleftarrow{\partial }_{x}-\overrightarrow{\partial }
_{x}\right) \psi ^{\prime }\left( X\right) dtd\mathbf{r}_{\bot }.
\label{norm1}
\end{eqnarray}
Note that for two solutions with different quantum numbers $n$, the inner
product $\left( \psi ,\psi ^{\prime }\right) _{x}$ can be easily calculated,
\begin{eqnarray}
&&\left( \psi _{n},\psi _{n^{\prime }}^{\prime }\right) _{x}=\mathcal{I\,}
(2\pi )^{d-1}\delta _{n,n^{\prime }},\   \notag \\
&&\mathcal{I}=\varphi _{n}^{\ast }\left( x\right) \left( i\overleftarrow{
\partial }_{x}-i\overrightarrow{\partial }_{x}\right) \varphi _{n}^{\prime
}\left( x\right) .  \label{norm2}
\end{eqnarray}
Solutions $_{\;\zeta }\psi _{n}\left( X\right) $ and $^{\;\zeta }\psi
_{n}\left( X\right) $ can be decomposed through each other as follows: 
\begin{eqnarray}
&&^{\;\zeta }\psi _{n}\left( X\right) =\,_{+}\psi _{n}(X)g\left(
_{+}\left\vert ^{\zeta }\right. \right) -\,_{-}\psi _{n}(X)g\left(
_{-}\left\vert ^{\zeta }\right. \right) ,  \notag \\
&&_{\;\zeta }\psi _{n}\left( X\right) =\,^{\;-}\psi _{n}\left( X\right)
g\left( ^{-}\left\vert _{\zeta }\right. \right) -\,^{\;+}\psi _{n}\left(
X\right) g\left( ^{+}\left\vert _{\zeta }\right. \right) ,  \label{e6}
\end{eqnarray}%
where the expansion coefficients are defined by the equations 
\begin{equation*}
\left( \ _{\zeta }\psi _{n},\ ^{\;\zeta ^{\prime }}\psi _{n^{\prime
}}\right) _{x}=g\left( _{\zeta }\left\vert ^{\zeta ^{\prime }}\right.
\right) \delta _{n,n^{\prime }},\quad g\left( ^{\zeta ^{\prime }}\left\vert
_{\zeta }\right. \right) =g\left( _{\zeta }\left\vert ^{\zeta ^{\prime
}}\right. \right) ^{\ast }.
\end{equation*}

Equation (\ref{e3}) can be written in the following form 
\begin{equation*}
\left[ \frac{\mathrm{d}^{2}}{\mathrm{d}\xi ^{2}}+\xi ^{2}-\lambda \right]
\varphi _{n}\left( x\right) =0,\quad \xi =\frac{eEx-p_{0}}{\sqrt{eE}},\quad
\lambda =\frac{\pi _{\bot }^{2}}{eE}.
\end{equation*}
Its general solution can be written in terms of an appropriate pair of
linearly independent Weber parabolic cylinder functions (WPCFs), either $
D_{\rho }[(1-\mathrm{i})\xi ]$ and $D_{-1-\rho }[(1+\mathrm{i})\xi ],$ or $
D_{\rho }[-(1-\mathrm{i})\xi ]$ and $D_{-1-\rho }[-(1+\mathrm{i})\xi ]$,
where $\rho =-\mathrm{i}\lambda/2-1/2$.

Using asymptotic expansions of WPCFs, one can construct the functions $
_{\;\zeta }\varphi _{n}\left( x\right) $ and $^{\;\zeta }\varphi _{n}\left(
x\right) $,  
\begin{eqnarray}
&&\ _{+}\varphi _{n}\left( x\right) =\ _{+}\mathcal{N}D_{-1-\rho }[-(1+
\mathrm{i})\xi ]\sim e^{-i\xi ^{2}/2},\;\;\xi \rightarrow -\infty ,\;p^{%
\mathrm{L}}=-\xi \sqrt{eE}  \notag \\
&&\ _{-}\varphi _{n}\left( x\right) =\ _{-}\mathcal{N}D_{\rho }[-(1-\mathrm{i
})\xi ]\sim e^{i\xi ^{2}/2},\;\;\xi \rightarrow -\infty ,\;p^{\mathrm{L}%
}=\xi \sqrt{eE};  \label{c1} \\
&&\ ^{+}\varphi _{n}\left( x\right) =\ ^{+}\mathcal{N}D_{\rho }[(1-\mathrm{i}
)\xi ]\sim e^{i\xi ^{2}/2},\;\;\xi \rightarrow \infty ,\;p^{\mathrm{R}}=\xi 
\sqrt{eE},  \notag \\
&&\ ^{-}\varphi _{n}\left( x\right) =\ ^{-}\mathcal{N}D_{-1-\rho }[(1+
\mathrm{i})\xi ]\sim e^{-i\xi ^{2}/2},\;\;\xi \rightarrow \infty ,\;p^{%
\mathrm{R}}=-\xi \sqrt{eE},  \notag \\
&&\ _{\zeta }\mathcal{N\ }\mathcal{=\ }^{\zeta }\mathcal{N}=\left(
2eE\right) ^{-1/4}e^{\pi \lambda /8}.  \label{c2}
\end{eqnarray}
Their \textrm{in-} and \textrm{out-}classifications are related with signs
of the asymptotic momenta $p^{\mathrm{L}}$ and $p^{\mathrm{R}}$ (see, Refs. 
\cite{L-field}). Namely, 
\begin{equation}
_{+}\psi _{n},\ ^{+}\psi _{n}\ \mathrm{are\ in-states,\ and\ }_{-}\psi
_{n},\ ^{-}\psi _{n}\mathrm{\ are}\text{ }\mathrm{out-states}.  \notag
\end{equation}

It is useful to construct two different complete sets of solutions of the
Klein-Gordon equation (\ref{1.1}) that are not stationary states and have
the following form: 
\begin{equation}
\psi _{\sigma }\left( X\right) =\varphi _{\sigma }\left( t,x\right) \varphi
_{\mathbf{p}_{\bot }}\left( \mathbf{r}_{\bot }\right) ,  \label{c5}
\end{equation}
where $\sigma$ is a set of quantum numbers, which will be defined below. In
this case the function $\varphi _{\sigma }\left( t,x\right)$ satisfies the
equation: 
\begin{equation}
\left\{ \hat{p}_{x}^{2}-\left[ \hat{p}_{0}-U\left( x\right) \right] ^{2}+ 
\mathbf{p}_{\bot }^{2}+m^{2}\right\} \varphi _{\sigma }\left( t,x\right)
=0,\quad \hat{p}_{0}=i\partial _{t}.  \label{c6}
\end{equation}
This equation admits integrals of motion in the class of linear differential
operators of the first order, which are: 
\begin{equation}
\hat{Y}_{0}=-ie,\quad \hat{Y}_{1}=\partial _{t},\quad \hat{Y}_{2}=\partial
_{x}+ieEt,\quad \hat{Y}_{3}=x\partial _{t}+t\partial _{x}+\frac{ieE}{2}
(t^{2}+x^{2}).  \notag
\end{equation}
The operators $\hat{Y}_{a}$, $a=0,1,2,3$ form a four-dimensional Lie algebra 
$\mathfrak{g}$ with nonzero commutation relations 
\begin{equation*}
\lbrack \hat{Y}_{1},\hat{Y}_{2}]=-E\,\hat{Y}_{0},\quad \lbrack \hat{Y}_{1}, 
\hat{Y}_{3}]=\hat{Y}_{2},\quad \ [\hat{Y}_{2},\hat{Y}_{3}]=\hat{Y}_{1}.
\end{equation*}

Equation (\ref{c6}) can be considered as an equation for the eigenfunctions
of the Casimir operator $K(-i\hat{Y})=-2E\,\hat{Y}_{0}\hat{Y}_{3}+ \hat{Y}
_{1}^{2}-\hat{Y}_{2}^{2}$, 
\begin{equation*}
-K(-i\hat{Y})\varphi _{\sigma }\left( t,x\right) =\left( \mathbf{p}_{\bot
}^{2}+m^{2}\right) \varphi _{\sigma }(t,x).
\end{equation*}

At this stage, we follow a non-commutative integration method of linear
differential equations \cite{nc1,nc2,nc3}, which allows us to construct a
complete set of solutions based on a symmetry of the equation. We define an
irreducible representation of the Lie algebra $\mathfrak{g}$ in the space of
functions of the variable $\tilde{p}\in (-\infty ,+\infty )$ by the help of
the operators $\ell _{a}(\tilde{p},\partial _{\tilde{p}},j)$, 
\begin{align*}
& \ell _{0}(\tilde{p},\partial _{\tilde{p}},j)=ie,\quad \ell _{1}(\tilde{p}
,\partial _{\tilde{p}},j)=-eE\partial _{\tilde{p}}+\frac{i}{2}\tilde{p}, \\
& \ell _{2}(\tilde{p},\partial _{\tilde{p}},j)=eE\partial _{\tilde{p}}+\frac{
i}{2}\tilde{p},\quad \ell _{3}(\tilde{p},\partial _{\tilde{p}},j)=-\tilde{p}
\partial _{\tilde{p}}+ij-\frac{1}{2},\quad j>0,  \notag
\end{align*}
where $j$ parameterizes the non-degenerate adjoint orbits of a Lie algebra $
\mathfrak{g}$. The following relations hold true: 
\begin{eqnarray*}
&&\ [\ell _{1},\ell _{2}]=-E\,\ell _{0},\quad \lbrack \ell _{1},\ell
_{3}]=\ell _{2},\quad \ [\ell _{2},\ell _{3}]=\ell _{1}, \\
&&\ K(-i\ell (\tilde{p},\partial _{\tilde{p}},j))=(2eE)j.
\end{eqnarray*}
Integrating the equations 
\begin{equation}
\left[ \hat{Y}_{a}+\ell _{a}(\tilde{p},\partial _{\tilde{p}},j)\right]
\varphi _{\sigma }\left( t,x\right) =0  \label{br09}
\end{equation}
together with equation (\ref{e3}), we fix $j=-\lambda /2$ and derive a set
of solutions which is characterized by quantum numbers $\sigma =(\tilde{p},
\mathbf{p}_{\bot })$, 
\begin{eqnarray}
&&_{-}^{+}\varphi _{\sigma }(t,x)=\,_{-}^{+}C_{\sigma }\exp \left( ie\frac{E 
}{2}\left[ \frac{1}{2}x_{-}^{2}-t^{2}\right] -\frac{i}{2}\left[ \lambda -i 
\right] \left( \ln \frac{\pm i\pi _{-}}{\sqrt{eE}}\right) -\frac{i}{2}\tilde{
p}\,x_{+}\right) ,  \notag \\
&&\pi _{-}=\tilde{p}+eEx_{-},\quad x_{\pm }=t\pm x.  \label{br01b}
\end{eqnarray}

The parameter $\tilde{p}$ is an eigenvalue of the symmetry operator $i(\hat{Y
}_{1}+\hat{Y}_{2})$: 
\begin{equation*}
i(\hat{Y}_{1}+\hat{Y}_{2})\,_{-}^{+}\varphi_{\sigma }(t,x)=\tilde{p}
\,_{-}^{+}\varphi_{\sigma }(t,x).
\end{equation*}

One can interpret the quantum numbers $\sigma$ from the point of view of the
orbit method: the parameter $\lambda =(m^{2}+\mathbf{p}_{\bot }^{2})/(eE)$
describes the Casimir operator $K(-i\hat{Y})$ spectrum and parameterizes the
non-degenerate orbits of the co-adjoint representation of the local Lie
group $\exp\mathfrak{g}$ (in our case, the orbits are hyperbolic paraboloids), 
and the region of variation of the parameter $\tilde{p}$ is a Lagrangian 
submanifold to these orbits.

In order to classify solutions (\ref{c5}), we define a direct and an inverse
integral transform that relate these solutions to solutions (\ref{e2}) that
are stationary states, eigenfunctions for the operator $\hat{p}_{0}$.

We represent solutions of both equation (\ref{e3}) and 
\begin{equation*}
\hat{p}_{0}\,\varphi _{n}^{\left( \pm \right) }(t,x)=p_{0}\,\varphi
_{n}^{\left( \pm \right) }(t,x)
\end{equation*}
in the following form 
\begin{equation}
\,\varphi _{n}^{\left( \pm \right) }(t,x)=(2\pi eE)^{-1/2}\int_{-\infty
}^{+\infty }M^{\ast }(p_{0},\tilde{p})\,_{-}^{+}\varphi _{\sigma }\left(
t,x\right) dp.  \label{br3.1}
\end{equation}
Taking into account condition (\ref{br09}) one gets an equation for the
function $M(p_{0},\tilde{p})$: 
\begin{equation*}
-i\ell _{1}(\tilde{p},\partial _{\tilde{p}},j)M(p_{0},\tilde{p}
)=p_{0}M(p_{0},\tilde{p}).
\end{equation*}
We choose its particular solution
\begin{equation}
M(p_{0},\tilde{p})=\exp \left( \frac{i}{4eE}[\tilde{p}^{2}-4\tilde{p}
\,p_{0}]\right) ,  \label{br3.4}
\end{equation}
which satisfies the orthogonality relation
\begin{equation}
\int_{-\infty }^{+\infty }M^{\ast }(p_{0},\tilde{p})M(p_{0},\tilde{p}
^{\prime })dp_{0}=2\pi eE\,\delta (\tilde{p}-\tilde{p}^{\prime }).
\label{br3.5}
\end{equation}
The inverse to (\ref{br3.1}) transform reads: 
\begin{equation}
_{-}^{+}\varphi _{\sigma }(t,x)=(2\pi eE)^{-1/2}\int_{-\infty }^{+\infty
}M(p_{0},\tilde{p})\varphi _{n}^{(\pm )}(t,x)dp_{0}.  \label{br3.6}
\end{equation}

Thus, we have defined direct (\ref{br3.1}) and inverse (\ref{br3.6})
integral transformation with kernel (\ref{br3.4}) that converts solutions 
(\ref{br01b}) to solutions that are eigenfunctions for the operator $\hat{p}
_{0}$. Applying one of the integral transformations to solutions 
(\ref{br01b}), we get: 
\begin{eqnarray}
&&\ \ \varphi _{n}^{(\pm )}(t,x)=(2\pi eE)^{-1/2}\int_{-\infty }^{+\infty
}M^{\ast }(p_{0},\tilde{p})_{-}^{+}\varphi _{\sigma }(t,x)d\,\tilde{p} 
\notag \\
&&\ =\,_{-}^{+}C_{\sigma }(1-i)^{\rho +1}e^{\frac{ip_{0}^{2}}{2eE}
}e^{-ip_{0}t}D_{\rho }\left[ \pm (1-i)\xi \right].  \label{br3.11}
\end{eqnarray}

Then, comparing Eq. (\ref{br3.11}) with Eqs. (\ref{c1})--(\ref{c2}), we
obtain the following correspondence 
\begin{eqnarray}
\varphi _{n}^{(+)}(t,x) &\sim &\ ^{+}\mathcal{N}^{\prime
}e^{-ip_{0}t}D_{\rho }[+(1-\mathrm{i})\xi ]=\,^{+}\varphi _{n}\left(
x\right) e^{-ip_{0}t},  \label{gav1} \\
~\varphi _{n}^{(-)}(t,x) &\sim &\ _{-}\mathcal{N}^{\prime
}e^{-ip_{0}t}D_{\rho }[-(1-\mathrm{i})\xi ]=\,_{-}\varphi _{n}\left(
x\right) e^{-ip_{0}t}.  \notag
\end{eqnarray}

Transformation (\ref{br3.6}) allows one to derive orthonormality relations
on the hyperplane $x=\mathrm{const}$ for scalar particles constructing with
the help of functions $_{-}^{+}\varphi _{\sigma }(t,x)$, 
\begin{equation}
\left( _{-}^{+}\psi _{\sigma },_{-}^{+}\psi _{\sigma ^{\prime }}\right)
_{x}=\pm \delta _{\sigma ,\sigma ^{\prime }},  \label{gav2a}
\end{equation}
where 
\begin{equation}
_{-}^{+}\psi _{\sigma }\left( X\right) =\,_{-}^{+}\varphi _{\sigma
}(t,x)\varphi _{\mathbf{p}_{\bot }}\left( \mathbf{r}_{\bot }\right),
\label{gav2b}
\end{equation}
and determine the normalizing factors $\,_{-}^{+}C_{\sigma }$, 
\begin{equation*}
\,_{-}^{+}C_{\sigma }=\frac{1}{\sqrt{4\pi eE}}e^{\pi \lambda /4}.
\end{equation*}
Thus, we obtain: 
\begin{equation*}
\left( _{-}\psi _{\sigma },^{+}\psi _{\sigma ^{\prime }}\right)
_{x}=\,g\left( _{-}\left\vert ^{+}\right. \right) \delta _{\sigma ,\sigma
^{\prime }},\quad g\left( _{-}\left\vert ^{+}\right. \right) =ie^{\pi
\lambda /2}.
\end{equation*}

We introduce now the following notation: 
\begin{eqnarray*}
&&\ \varphi _{n}^{\left( \pm \right) }(t,x)=\,_{-}^{+}\varphi _{n}(x)\exp
\left( -ip_{0}t\right) , \\
&&\ _{-}^{+}\psi _{n}(X)=\varphi _{n}^{\left( \pm \right) }(t,x)\varphi _{ 
\mathbf{p}_{\bot }}\left( \mathbf{r}_{\bot }\right).
\end{eqnarray*}
It follows from (\ref{br3.1}) and (\ref{br3.6}): 
\begin{eqnarray}
&&\,_{-}^{+}\psi _{\sigma }\left( X\right) =(2\pi eE)^{-1/2}\int_{-\infty
}^{+\infty }M(p_{0},\tilde{p})\,_{-}^{+}\psi _{n}(X)dp_{0},  \label{gav3} \\
&&\,_{-}^{+}\psi _{n}(X)=(2\pi eE)^{-1/2}\int_{-\infty }^{+\infty }M^{\ast
}(p_{0},\tilde{p})\text{\thinspace }_{-}^{+}\psi _{\sigma }\left( X\right) d
\tilde{p}.  \notag
\end{eqnarray}

Let us consider another type of solutions, 
\begin{eqnarray}
&&_{+}\varphi _{\sigma }(t,x)=\theta \left( -\pi _{-}\right) \,^{+}\varphi
_{\sigma }(t,x),  \label{br01c} \\
&&^{-}\varphi _{\sigma }(t,x)=\theta \left( +\pi _{-}\right) \,_{-}\varphi
_{\sigma }(t,x).  \notag
\end{eqnarray}
The corresponding integral transformation is: 
\begin{eqnarray}
&&_{+}\varphi _{n}(t,x)=(2\pi eE)^{-1/2}\int_{-\infty }^{+\infty }M^{\ast
}(p_{0},\tilde{p})\theta \left( -\pi _{-}\right) \,^{+}\varphi _{\sigma
}(t,x)d\tilde{p}  \notag \\
&&\ =-\,^{+}C_{\sigma }\sqrt{\frac{2}{\pi }}(-[1+i])^{\rho -1}\Gamma (\rho
+1)e^{\frac{ip_{0}^{2}}{2eE}}e^{-ip_{0}t}D_{-\rho -1}\left[ -(1+i)\xi \right]
,  \notag \\
&&^{-}\varphi _{n}(t,x)=(2\pi eE)^{-1/2}\int_{-\infty }^{+\infty }M^{\ast
}(p_{0},\tilde{p})\theta \left( \pi _{-}\right) \,_{-}\varphi _{\sigma
}(t,x)d\tilde{p}  \notag \\
&&\ =-\,_{-}C_{\sigma }\sqrt{\frac{2}{\pi }}(-[1+i])^{\rho -1}\Gamma (\rho
+1)e^{\frac{ip_{0}^{2}}{2eE}}e^{-ip_{0}t}D_{-\rho -1}\left[ (1+i)\xi \right]
\label{br3.11g}
\end{eqnarray}
such that 
\begin{eqnarray}
&&_{+}\varphi _{n}(t,x)\sim \,_{+}\mathcal{N}^{\prime
}e^{-ip_{0}t}D_{-1-\rho }[-(1+\mathrm{i})\xi ]=\;_{\;+}\varphi _{n}\left(
x\right) e^{-ip_{0}t},  \notag \\
&&^{-}\varphi _{n}(t,x)\sim \ ^{-}\mathcal{N}^{\prime
}e^{-ip_{0}t}D_{-1-\rho }[(1+\mathrm{i})\xi ]=\;^{\;-}\varphi _{n}\left(
x\right) e^{-ip_{0}t}.  \label{gav1b}
\end{eqnarray}

Using functions $\,_{+}^{-}\varphi _{\sigma }(t,x)$, we construct a new set
of solutions: 
\begin{equation}
\,_{+}^{-}\psi _{\sigma }(t,x)\left( X\right) =\sqrt{e^{\pi \lambda }-1}
\,_{+}^{-}\varphi _{\sigma }(t,x)\varphi _{\mathbf{p}_{\bot }}\left( \mathbf{
\ r}_{\bot }\right),  \label{gav6}
\end{equation}
which satisfies the following orthonormality relations: 
\begin{eqnarray*}
&&\left( _{+}\psi _{\sigma },_{-}\psi _{\sigma ^{\prime }}\right)
_{x}=0,\;\left( \,^{-}\psi _{\sigma },^{+}\psi _{\sigma ^{\prime }}\right)
_{x}=0,  \notag \\
&&(_{+}^{-}\psi _{\sigma },_{+}^{-}\psi _{\sigma })_{x}=\pm \delta _{\sigma
,\sigma ^{\prime }}.
\end{eqnarray*}
From (\ref{br3.11g}) and (\ref{br3.5}) follows the integral transformations 
\begin{eqnarray}
&&\,_{+}^{-}\psi _{\sigma }\left( X\right) =(2\pi eE)^{-1/2}\int_{-\infty
}^{+\infty }M(p_{0},\tilde{p})\,_{+}^{-}\psi _{n}(X)dp_{0},  \notag \\
&&\,_{+}^{-}\psi _{n}(X)=(2\pi eE)^{-1/2}\int_{-\infty }^{+\infty }M^{\ast
}(p_{0},\tilde{p})\text{\thinspace }_{+}^{-}\psi _{\sigma }\left( X\right)
d\,\tilde{p}.  \label{brTr2}
\end{eqnarray}

There exist useful relations between solutions $_{\zeta }\psi _{\sigma
}\left( X\right)$ and $^{\zeta }\psi _{\sigma }\left( X\right)$. Each of
them is complete for a given $\sigma$ and can be decomposed through each
other as follows: 
\begin{eqnarray}
&&^{\;\zeta }\psi _{\sigma }\left( X\right) =\,_{+}\psi _{\sigma }(X)g\left(
_{+}\left\vert ^{\zeta }\right. \right) -\,_{-}\psi _{\sigma }(X)g\left(
_{-}\left\vert ^{\zeta }\right. \right) ,  \notag \\
&&_{\;\zeta }\psi _{\sigma }\left( X\right) =\,^{\;-}\psi _{\sigma }\left(
X\right) g\left( ^{-}\left\vert _{\zeta }\right. \right) -\,^{\;+}\psi
_{\sigma }\left( X\right) g\left( ^{+}\left\vert _{\zeta }\right. \right),
\label{br8}
\end{eqnarray}
Equations 
\begin{equation*}
\left( \ _{\zeta }\psi _{\sigma },\ ^{\;\zeta ^{\prime }}\psi _{\sigma
^{\prime }} \right) _{x}=g\left( _{\zeta }\left\vert ^{\zeta ^{\prime
}}\right. \right) \delta _{\sigma ,\sigma ^{\prime }},\quad g\left( ^{\zeta
^{\prime }}\left\vert _{\zeta }\right. \right) =g\left( _{\zeta }\left\vert
^{\zeta ^{\prime }}\right. \right) ^{\ast }.
\end{equation*}
allow us to calculate coefficients $g\left( _{-}\left\vert ^{-}\right.
\right)$ and $g\left( _{+}\left\vert ^{+}\right. \right)$, 
\begin{equation*}
g\left( ^{-}\left\vert _{-}\right. \right) =-\sqrt{e^{\pi \lambda }-1},\quad
g\left( ^{+}\left\vert _{+}\right. \right) =+\sqrt{e^{\pi \lambda }-1}.
\end{equation*}
We note that the relations (\ref{br8})\ are similar to relations (\ref{e6})
that were established for the solutions $_{\zeta }\psi _{n}\left( X\right)$
and $^{\zeta }\psi _{n}\left( X\right)$ (in this case the coefficients $g$
do not depend on $p_{0}$ and $\tilde{p}$). From (\ref{gav1}) and 
(\ref{gav1b}) it follows 
\begin{equation}
_{+}\psi _{\sigma },\,^{+}\psi _{\sigma }\ \mathrm{are\ in-states,\ and\ }
_{-}\psi _{\sigma },\,^{-}\psi _{\sigma }\mathrm{\ are}\text{ }\mathrm{\
out-states}.  \notag
\end{equation}

From Eq. (\ref{br01c}) it follows 
\begin{eqnarray}
\ _{+}\psi _{\sigma }(X) &=&0,\quad \pi _{-}>0,  \notag \\
\ ^{-}\psi _{\sigma }(X) &=&0, \quad \pi _{-}<0.  \label{br9}
\end{eqnarray}
Then, taking into account equations (\ref{br8}) and (\ref{br9}), we get: 
\begin{eqnarray}
&&_{+}\psi _{\sigma }\left( X\right) =g\left( _{+}\left\vert ^{-}\right.
\right) ^{-1}\left[ _{-}\psi _{\sigma }\left( X\right) g\left(
_{-}\left\vert ^{-}\right. \right) +\,^{-}\psi _{\sigma }\left( X\right) 
\right] =0,\quad \pi _{-}>0,  \notag \\
&&^{-}\psi _{\sigma }(X)=g\left( ^{-}\left\vert _{+}\right. \right) ^{-1} 
\left[ \ ^{+}\psi _{\sigma }\left( X\right) g\left( ^{+}\left\vert
_{+}\right. \right) +\ _{+}\psi _{\sigma }\left( X\right) \right] =0,\quad
\pi _{-}<0.  \label{gav9}
\end{eqnarray}
Since the coefficient $g\left( _{+}\left\vert ^{-}\right. \right)$ is not
zero for all $\sigma$, the equations (\ref{gav9}) imply a direct connection
between the solutions $_{\zeta }\psi _{\sigma }\left( X\right)$ and $^{\zeta
}\psi _{\sigma }\left( X\right) $ normalized on the hyperplane $x=\mathrm{
const}$, 
\begin{equation*}
\ ^{-}\psi _{\sigma }\left( X\right) =-\,_{-}\psi _{\sigma }\left( X\right)
g\left( _{-}\left\vert ^{-}\right. \right) \theta \left( \pi _{-}\right)
,\;\ _{+}\psi _{\sigma }\left( X\right) =-\ ^{+}\psi _{\sigma }\left(
X\right) g\left( ^{+}\left\vert _{+}\right. \right) \theta \left( -\pi
_{-}\right).
\end{equation*}

Thus, using the noncommutative integration method for equation (\ref{c6}) we
obtained \textrm{in}- and \textrm{out}-states of scalar particles in terms
of new solutions (\ref{gav2b}) and (\ref{gav6}), which are non-stationary
and are determined by a set of quantum numbers $\sigma$. Solutions $
\{_{+}\psi _{\sigma },\,^{+}\psi _{\sigma }\}$ describe \textrm{in}
-solutions, and solutions $\{_{-}\psi _{\sigma },\,^{-}\psi_{\sigma }\}$
describe \textrm{out}-states. Using integral transformations (\ref{gav3})
and (\ref{brTr2}) the solutions $_{\zeta }\psi _{\sigma }\left(X\right)$ and 
$^{\zeta }\psi _{\sigma }\left( X\right)$ are related to the well-known
stationary solutions $\ _{\zeta }\psi _{n}\left( X\right)$ and $\ ^{\zeta
}\psi _{n}\left( X\right)$ (see Ref. \cite{L-field}).

\section{Acknowledgement}

Authors acknowledge support from Tomsk State University Competitiveness
Improvement Program and the partial support from the Russian Foundation for
Basic Research (RFBR), under the project No. 18-02-00149; Gitman is also
supported by the Grant No. 2016/03319-6, Funda\c{c}\~ao de Amparo \`{a}
Pesquisa do Estado de S\~{a}o Paulo (FAPESP) and permanently by Conselho
Nacional de Desenvolvimento Cient\'ifico e Tecnol\'ogico (CNPq).

\end{document}